\RequirePackage{fix-cm}

\documentclass[smallextended]{svjour3}       
\smartqed  
\usepackage{booktabs}
\usepackage{graphicx}
\usepackage{xcolor}
\usepackage{fancybox}
\usepackage{pdflscape}
\usepackage{caption}
\usepackage{subcaption}
\usepackage{tabularx}
\usepackage{longtable}
\usepackage{array}
\usepackage{fontawesome}
\usepackage{multirow}
\usepackage{array}
\usepackage{easyReview}
\usepackage{soul}
\usepackage{fancyhdr}

\usepackage{subcaption}
\usepackage{booktabs}
\usepackage{setspace}
\usepackage{tikz}
\usepackage{amssymb}
\usepackage{booktabs}
\usepackage{enumitem}
\usepackage{changepage}
\usepackage{hyperref}
\hypersetup{
    colorlinks=true,
    linkcolor=black,
    filecolor=black,      
    urlcolor=black,
    citecolor=black,
}
\usepackage{graphicx}

\usepackage{fancybox}
\usepackage{longtable}
\usepackage{tasks}
\usepackage{pdfpages}

\usepackage{type1cm}        
\usepackage{makeidx}         
\usepackage{graphicx}        
\usepackage{multicol}        
\usepackage[bottom]{footmisc}
\usepackage{newtxtext}       
\usepackage{newtxmath}       

\makeindex              
\usepackage{cite}

\newcommand\BibTeX{{\rmfamily B\kern-.05em \textsc{i\kern-.025em b}\kern-.08em
T\kern-.1667em\lower.7ex\hbox{E}\kern-.125emX}}

\usepackage{booktabs}

\definecolor{principle01}{RGB}{143, 170, 220}
\definecolor{principle02}{RGB}{169, 209, 142}
\definecolor{principle03}{RGB}{244, 177, 131}
\definecolor{principle04}{RGB}{191, 191, 191}

\usepackage{lipsum} 
\usepackage{xargs}

\newcommand{\manuscript}{{paper}}

\newcommand{\quan}{quaN}
\newcommand{\qual}{quaL}

\newcommand{\edit}[1]{\textcolor{black}{#1}}

\begin{document}
\title{\edit{Guiding Principles  for Mixed Methods Research in \\ Software Engineering}}

\titlerunning{Guiding Principles for Mixed Methods Research in SE}        

\author{Margaret-Anne Storey         \and
        Rashina Hoda      \and
        Alessandra Maciel Paz Milani       \and
       \\  Maria Teresa Baldassarre
}
\authorrunning{Storey et al.} 

 \institute{Margaret-Anne Storey \at
               University of Victoria \\
               Victoria, BC, 
              Canada\\
               \email{mstorey@uvic.ca}           
            \and 
            Rashina Hoda\at
              Monash University \\
               Melbourne, VIC,
              Australia\\
               \email{rashina.hoda@monash.edu} 
            \and
            Alessandra Maciel Paz Milani \at
               University of Victoria \\
               Victoria, BC, 
              Canada\\
               \email{amilani@uvic.ca} 
                \and
            Maria Teresa Baldassarre \at
               University of Bari \\
               Bari, 
              Italy\\
               \email{mariateresa.baldassarre@uniba.it} 
}


\maketitle

\begin{abstract}

\textit{Mixed methods research is often used in software engineering, but researchers outside of the social or human sciences often lack experience when using these designs. This paper provides \edit{guiding principles} and advice on how to design mixed method research, and to encourage the intentional, rigorous, and innovative application of mixed methods in software engineering. It also presents key properties of core mixed method research designs. 
Through a number of fictitious but recognizable software engineering research scenarios, we showcase how to choose suitable mixed method designs and consider the inevitable trade-offs any design choice leads to. We describe several antipatterns that illustrate what to avoid in mixed method research, and when mixed method research should be considered over other approaches.} 


\end{abstract}

\section{Introduction}
\label{sec:intro}

Software engineering (SE) researchers use a variety of research methods, often mixing and matching them. Mixing methods is particularly pertinent in today's complex socio-technical SE contexts~\cite{storey_who_2020, hoda2024Book} where quantitative and qualitative methods, when combined, can bring insights on the nuanced nature of the problem and/or solution being studied. \edit{Before discussing the use and state of mixed methods in SE research, we first share the definition of mixed methods research we use throughout our paper\footnote{This definition is in line with the definitions by other scholars of mixed methods research, such as ~\cite{poth_sage_2023}.}: 
}

\begin{center}
\noindent\fbox{%
    \parbox{32.5em}{
    \textbf{Mixed methods research (MMR)} is \textit{a research approach where multiple methods are used to collect, analyze, and integrate both qualitative and quantitative data to address a research problem and produce novel insights.}}
}
\end{center}
\vspace{0.2cm}

There is widespread prevalence of mixing methods in the \edit{social sciences~\cite{bryman_integrating_2006, creswell_best_2011, poth_innovation_2018, ladner_mixed_2019,poth_sage_2023}} \edit{ and well developed guidelines for mixed methods research in the related discipline of Information Systems (IS)
~\cite{venkatesh2013bridging, venkatesh2016guidelines}}.
However, while mixed methods are recognized and utilized in Software Engineering (SE), there is limited discussion on the unique methodological issues, challenges, and emerging trends shaping mixed methods research in this domain. Di Penta and Tamburri~\cite{Dipenta} explore the integration of quantitative and qualitative approaches in empirical SE, highlighting how each can complement the other. However, there remains a gap in providing actionable guidance for applying mixed methods research specifically within SE. This paper aims to fill that gap by offering \edit{practical guiding principles} to support researchers in designing studies that effectively and rigorously combine qualitative (quaL) and quantitative (quaN) methods in an SE context.

Our paper is aimed at researchers unfamiliar with mixed methods and those keen to improve their understanding and practice.  We share key concepts and provide guidance on how to mix and match methods, presenting different properties of mixed methods and some common designs. 
The timing of this paper is important, as SE researchers today face a complex world where software is used to address complex problems with wide-scale societal impact. There is also a strong drive for innovation in the face of rapidly changing technologies and practices, all occurring in socio-technical contexts~\cite{storey_who_2020, hoda_socio-technical_2022}. 
The increasing interest in studying the adoption of generative AI technologies in everyday SE practice is but one latest example of the constantly changing SE research landscape \cite{bano2023LLM}. Choosing which mixed methods to use and knowing how to use them effectively is challenging, and is often overlooked in SE research and education  curricula and thus, experience is lacking in how to use these methodological designs. 

Inspired by a previous paper one of us co-authored on selecting  empirical methods for SE research~\cite{easterbrook_selecting_2008}, this paper includes practical illustrative examples in the form of four fictional SE researchers and research scenarios, representing common mixed method contexts, design decisions, and outcomes. 
These four research scenarios are used to exemplify and motivate the use of mixed method research designs throughout the \manuscript.
While not exhaustive, these scenarios not only help to guide design choice but also emphasize the trade-offs faced in making some common mixed method study design decisions. As well as showcasing design choices for a variety of scenarios, the paper also calls out best practices and a number of antipatterns that should be avoided. 

\edit{Through this paper, we aim to demystify, clarify, and define MMR and its various properties, designs, and application scenarios for the SE research community. We encourage and call for SE researchers to use mixed methods and to do so in a more principled way. Our hope is that it will serve as the start of a robust conversation in the community about intentional, rigorous, and innovative use of MMR in SE, one that others inevitably will build upon in the future.}

The rest of this \manuscript ~is organized as follows. 
In Section~\ref{sec:background}, we present a brief background of the most prominent and relevant influences in mixed methods research design over the decades.
In Section~\ref{sec:principles}, we present four core \edit{principles} that underlie mixed method research: \textit{methodological rationale}, \textit{novel integrated insights}, \textit{procedural rigor} and \textit{ethical research}. 
In Section~\ref{sec:landscape}, we discuss some notable properties of mixed method designs, as well as common mixed methods research designs.
In Section~\ref{sec:scenarios}, we present four scenarios of research studies that fictional researchers pursue. We describe the methodological choices these characters make and how they followed the four core principles described in Section~\ref{sec:principles}. In Section~\ref{sec:antipatterns}, we use the concept of antipatterns to present what not to do in mixed method research. In Section~\ref{sec:discussion}, we discuss mixed methods research designs with other research designs (such as multi-method, case studies and secondary studies) and share a call to action before concluding the paper in Section~\ref{sec:conclusions}. This \manuscript~includes an Appendix which contains a glossary of  terms we use throughout the paper for ease of reference.

\section{A Brief Overview of Mixed Methods Research} 
\label{sec:background}

Early researchers recognized that many research questions can be addressed using both qualitative (\qual) and quantitative (\quan) methods that are combined in various stages of the investigation process, from data collection and analysis, to synthesis and to the reporting of findings.
The need to integrate \qual~descriptions and \quan~measurements to understand phenomena is not new. Early roots of mixed methods date back as far as 1000 BC in the natural sciences and were used by Babylonian astronomers. Aristotle and Galileo also referred to use of mixed methods for classifying animals and observing planets respectively~\cite{maxwell_expanding_2016}.

It is only in the mid-20th century, with roots in disciplines such as sociology and anthropology, that mixed methods research (MMR) gained traction with the formalization of approaches by authors including Creswell and Plano Clark~\cite{creswell_designing_2018}. 
According to these authors, the early development of mixed methods began in the 1950s and extended through the 1980s, marking the initial interest in incorporating multiple methods within a single study. These authors did not explicitly identify these approaches as \textit{mixed methods}. Furthermore, the different typologies of mixed method designs did not emerge until the 1990s~\cite{maxwell_expanding_2016,Greene1989}.  
In the social sciences field, Campbell and Fiske's paper from 1959~\cite{campbell_convergent_1959} is recognized as the first adoption of multiple (\edit{that included mixed}) research methods for validation purposes. They defined \edit{the multiple method} design as \textit{multi-operationalism} since the authors propose a multitrait, multimethod matrix that intersects \qual-\quan~methods with a specific set of traits as a means for enforcing validity of research conclusions. 
In the mid 1960s, Webb \emph{et al.}~\cite{webb_unobtrusive_1966} introduced the concept of \textit{triangulation}, followed by Denzin in the 1970s~\cite{denzin_research_1978} who detailed how to triangulate using both within-methods triangulation and between-methods triangulation, emphasizing the latter as a more powerful means to overcome bias and as a way to pragmatically mix methods. 

Although the combination of \qual-\quan~methods has been used in practice for many years, the first evidence of a more formal use of mixed methods is identifiable only in the late 1980s \edit{\cite{Kuhn1977}, as discussed by Guest and Fleming~\cite{guest_mixed_2015} and Johnson \emph{et al.}~\cite{Johnson2004},\cite{johnson_toward_2007} who point out how mixed methods research combines qualitative and quantitative approaches, forming a \textit{third paradigm} that complements traditional methods to yield more comprehensive, balanced, and insightful results.}  
The common denominator in these early conceptualizations of mixed methods is that MMR combines \qual-\quan~elements. It is only recently that researchers have turned their attention to proposing \textit{typologies} of mixed methods which describe how to structure and design an MMR by combining \qual~and \quan~elements according to some criteria as stated by Teddlie and Tashakkori~\cite{teddlie_foundations_2009}. 

This rise of formalized mixed method approaches has led to numerous discussions, but often with different design representations and  overlapping terms. Indeed, not even the basic definition of what can be qualified as an MMR study sees consensus (see Section 1 in ~\cite{poth_sage_2023}).
Part of this issue originates in the ``paradigm war''~\cite{Bryman2008}
where Bryman discusses the historical association of \quan~research with positivism and \qual~research with constructivism. He argues that these conflicts have diminished over time, giving way to a more pragmatic approach that embraces methodological pluralism as researchers are increasingly combining quantitative and qualitative methods to enrich their studies, moving beyond rigid adherence to a single paradigm. This shift reflects a growing acceptance of mixed methods as a valuable research strategy. It is not our goal to review these research paradigms, design typologies or resolve paradigm debates in this paper, but it is essential to note that some research paradigms that underpin MMR, such as pragmatism, emphasize the importance of using the most effective methods to answer research questions. Moreover, the philosophical foundation of a researcher's choice will inevitably impact how \qual~and \quan~approaches are integrated. 

Today, MMR is widely recognized and utilized across diverse fields such as psychology, education, public health, \edit{information systems} and management, and these approaches offer a flexible and robust framework for addressing multifaceted research inquiries and advancing knowledge in complex socio-technical domains~\cite{poth_innovation_2018, ladner_mixed_2019}. \edit{In software engineering, we can learn much from related disciplines, such as information systems~\cite{venkatesh2013bridging, venkatesh2016guidelines}, that articulate principles and guidelines of how to conduct mixed methods research. The guidelines by Venkatesh \emph{et al.}~\cite{venkatesh2013bridging}, building on top of the mixed methods research by Tashakkori and Creswel~\cite{TashakkoriCreswell2008}, Teddlie and Tashakkori~\cite{teddlie2003major,teddlie_foundations_2009}, Creswell~\cite{creswell_research_2003} among others, clearly state that the choice to conduct mixed methods research should depend on the research question, purpose, and context.}   

\begin{figure}[]
    \centering
    \includegraphics[scale=0.162]{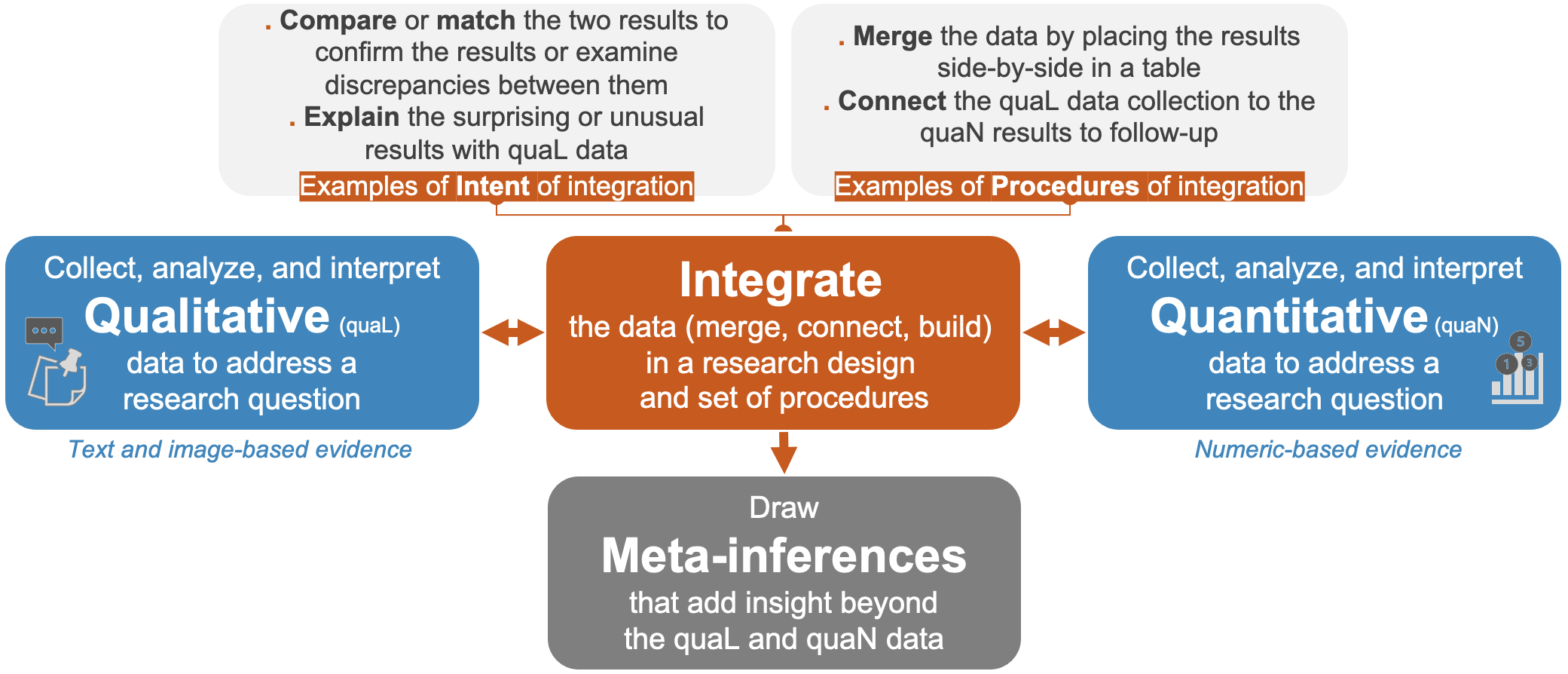}
    \caption{Interconnection of the core components for MMR: collecting, analyzing and interpreting (1) qualitative and (2) quantitative data to address a research question;  (3) integrating the data (merging, connecting, and building) in a research design and set of procedures; and (4) drawing meta-inferences that add new insight beyond the \qual~and \quan~data. Adapted from Creswell and Plano Clark~\cite[Chapter 2] {poth_sage_2023}.}    
    \label{fig:interconnection}
\end{figure}

With the increasing complexity of MMR designs, the problem of discussing and visually presenting complex designs to readers became apparent. Consequently, as noted by Creswell and Plano Clark~\cite[Chapter 2]{poth_sage_2023} when they revisited MMR designs twenty years after their important chapter was first published in 2003~\cite{tashakkori_handbook_2003}, the most recent studies decreased their focus on the \quan~and \qual~components and increased attention towards the integration processes across design types.
To reflect this change about the central elements of the MMR designs, Figure \ref{fig:interconnection} represents the simplification of the interconnection of the core components of an MMR design. 
From this perspective, treating the data beyond \qual~and \quan~analysis and gaining insight through integration and meta-inferences\footnote{See Appendix for a glossary of terms used through the paper. Integration is discussed in detail later in the paper.} become central. 

To conclude our brief overview of MMR, current trends in MMR can be listed as a call for openness and creativity during the research process, integrating mixed methods in interdisciplinary research, and the use of digital tools for data collection and analysis. %
Furthermore, researchers will be increasingly challenged to keep pace with a rapid evolution in design terminology and naming practices. 
Given such dynamism, paying attention to the many sources of contextual influences on the design of MMR becomes paramount~\cite{poth_sage_2023}. 
Recognizing the challenges these trends impose on researchers, we provide practical design guidance of MMR for SE researchers in our paper. \edit{In the next section, we describe four fundamental principles to be followed when using MMR.}

\section{Principles to Guide Mixed Method Research}
\label{sec:principles}

\begin{figure}[th]
    \centering
    \includegraphics[scale=0.15]{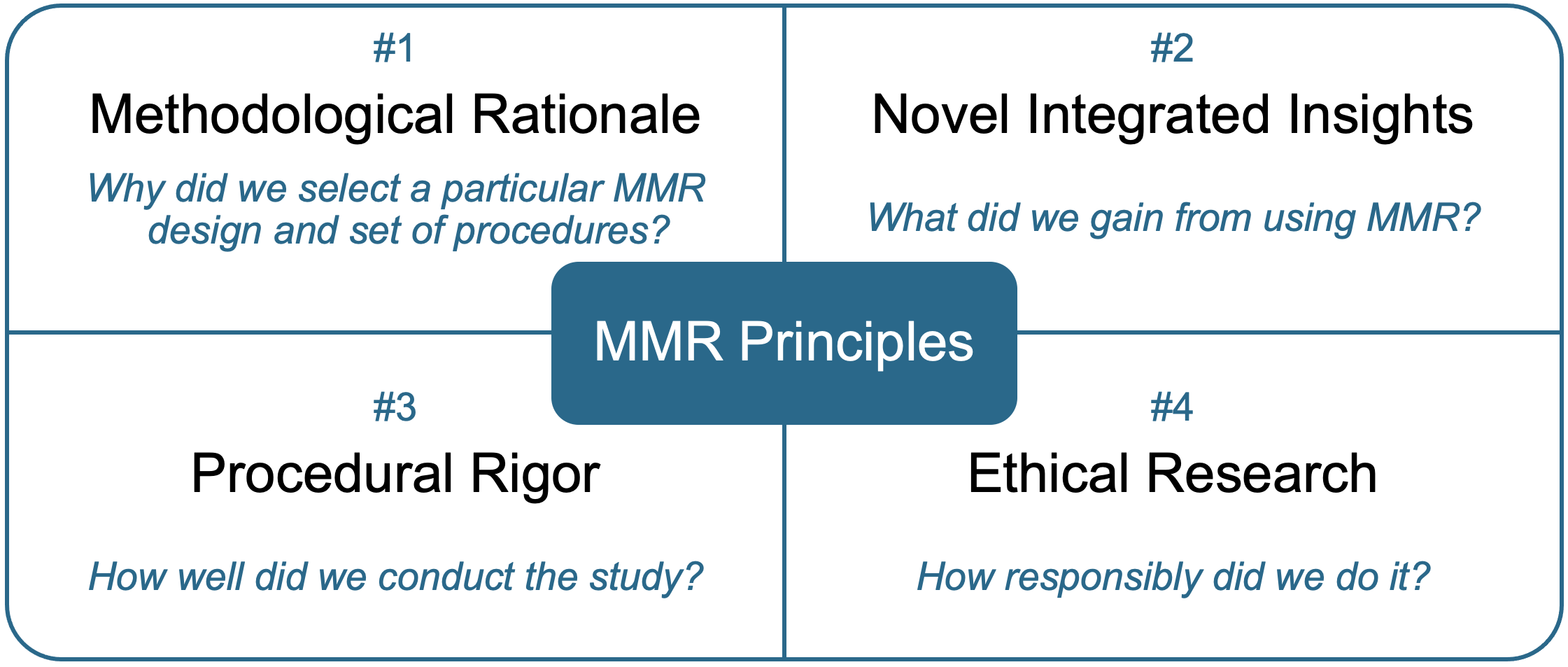}
    \caption{\edit{Four Guiding} Principles for Mixed Method Research \edit{in Software Engineering.}} 
    \label{fig:principles}
\end{figure}

\edit{Building on the references discussed in Section 2, along with the MMR characteristics highlighted by Poth~\cite{poth_innovation_2018}, Ladner~\cite{ladner_mixed_2019}, Creswell and Plano Clark~\cite{creswell_designing_2018}, we synthesize four guiding principles for conducting MMR in Software Engineering (see Figure~\ref{fig:principles}).
Our guiding principles are oriented by a pragmatic approach focused on describing key elements and properties of MMR rather than emphasizing the methodologies or philosophical foundations.}

Two of the principles presented in our paper, \textbf{methodological rationale} and development of \textbf{novel integrated insights} (meta-inferences), relate to understanding \textit{why} we should consider using an MMR design. The other two, \textbf{procedural rigor} and \textbf{ethical research}, relate to understanding \textit{how} we should go about designing and applying MMR. While some aspects of these principles apply to single method applications and can be considered good research practice in general, others are particularly relevant to MMR and generally more complex in that context. We delve into these four principles below.

\subsection{Methodological Rationale}

\begin{center}
\noindent\fbox{%
    \parbox{32.5em}
    {\textbf{Principle of Methodological Rationale}: Mixed method research study designs should explain how the mixed research design and data methods are relevant for addressing a given research problem. By adhering to this principle, an MMR study description can address the question \textit{why did we select a particular MMR design and set of procedures?}
}}
\end{center}

There are several conceptual \textbf{reasons} for using mixed methods.\footnote{The reasons we list are inspired by or earlier proposed by Ladner~\cite{ladner_mixed_2019}, 
Bryman~\cite{bryman_research_2007}, Creswell and Plano Clark~\cite{creswell_designing_2018} and McGrath~\cite{mcgrath_methodology_1995}.}  
The rationale for the methodology may be one or more of the following: 

\begin{itemize}
        \item \textbf{Complementarity:} one method helps to deepen, confirm or enhance the insights of another method~\cite{ladner_mixed_2019,bryman_social_2012}.
        \item \textbf{Expansion:} one method is used to ask different questions that are apparent up front or emerge throughout the study (in parallel or sequentially)~\cite{ladner_mixed_2019,bryman_social_2012}. 
        \item \textbf{Development:} one method is used to inform or improve another method (e.g., interviews may be used to inform the development of a questionnaire)~\cite{ladner_mixed_2019,bryman_social_2012}. 
        \item \textbf{Triangulation:} a different type of method is used to corroborate earlier insights leveraging multiple quality criteria afforded by use of different methods~\cite{mcgrath_methodology_1995} notably generalizability, realism, precision, and control. 
        \item \textbf{Credibility:} another method may be used to increase the integrity and truth of the findings from a former method~\cite{bryman_social_2012}. Mixed methods can increase credibility by bringing multi-dimensional philosophical and theoretical lenses to the findings~\cite{mertens_mixed_2013}.  
        \item \textbf{Explanation:} one method can be used to resolve previous contradictory, surprising or inconclusive findings~\cite{bryman_integrating_2006}. 
         \item \textbf{Increased design flexibility:} mixed methods may allow researchers to consider research design alternatives to offset limitations of a single method when faced with unexpected issues or unsatisfactory data. For example, many researchers that were planning to conduct ethnographic studies during the global pandemic pivoted to using alternative designs, such as a combination of online interviews, self-reported work diary entries, and data mining.\footnote{\url{https://www.microsoft.com/en-us/research/project/the-new-future-of-work/}} 
\end{itemize}

\subsection{Novel Integrated Insights}

\begin{center}
\noindent\fbox{%
    \parbox{32.5em}{
\textbf{Principle of Novel Integrated Insights}: Mixed methods research studies should generate new insights through the integration of the methods, which in turn addresses the reason(s) for the use of mixed methods. An MMR study report should clarify the novelty of the study findings and explain how methods and/or data were integrated. By adhering to this principle, an MMR study report can address the question \textit{what did we gain from using MMR?}
}}
\end{center}

An MMR study should deliver on its promise of additional benefits to what could have been afforded by using a single method~\cite{poth_innovation_2018}. 
\edit{Typically, once data analysis for each \qual/\quan~method is complete, the researcher compiles the results and findings (a phase often referred to as the inferential stage \cite{teddlie_foundations_2009}). In an MMR study, inferences are also drawn from both \qual~and \quan~analyses. However, to achieve a truly integrated approach, these conclusions should be synthesized into meta-inferences (see Fig.~\ref{fig:interconnection}), which may involve generating additional insights or linking findings from both methods to broader theories or implications.\footnote{Venkatesh \emph{et al.}~\cite{venkatesh2013bridging} provide an in-depth discussion on development and assessment of the quality of meta-inferences.}}
An MMR study may bring novel insights in terms of one or more of the following:

\begin{itemize}
    \item \textbf{Improved problem understanding:} a better understanding of the problem space and a fuller picture of the research question before rushing into a solution. 
    \item \textbf{Greater depth and breadth:} more nuanced and comprehensive coverage of the problem and solution spaces, with room for evaluating the proposed solutions.
    \item \textbf{Explaining unexpected results:} where the use of one method produced unexpected results, another can be employed in a bid to further explore why those unexpected results were encountered. 
    \item \textbf{Complementary storytelling:} \quan~data provides a numerical strength and statistical measure, and \qual~data provides richer insights into the experiences and perspectives of those close to the phenomenon. Together they help narrate a better and fuller story, with multiple ways to present findings in engaging formats.\footnote{Bryman refers to this as \textit{illustration}, or the idea of ``putting `meat on the bones' of `dry' \quan~findings''~\cite{bryman_social_2012}} 
\end{itemize}

\subsection{Procedural Rigor}

\begin{center}
\noindent\fbox{%

\parbox{32.5em}{
\textbf{Principle of Procedural Rigor}: Mixed method studies should apply the study methods rigorously and explain them effectively. By adhering to this principle, an MMR study report can address the question \textit{how well did we conduct the study?}
}}
\end{center}

Rigor in any research study is always a critical concern, and it is often a difficult concept for researchers 
to know how to assess. Rigor is particularly challenging for mixed methods.  MMR studies should first adhere to the rigor of the individual studies, but there are also some concerns that apply to the ``mixing'' aspect. 
For this principle, we are guided by a mixed methods rigor appraisal tool called MMAT (Mixed Methods Appraisal Tool)~\cite{oliveira_mixed_2021}. The following are posed as questions to be considered when self-reflecting on the rigor of a mixed design:   
\begin{itemize}
    \item \textbf{Is the use of mixed methods justified?} As mentioned above, methodological rationale (e.g., those listed above) should be clearly described. 
    \item \textbf{Are the methods effectively integrated to answer the research question?} It is important to appraise how well the data from the multiple methods are integrated to answer the research question posed and to appraise at what stage the methods are integrated: during collection, during analysis or during both collection and analysis.  
    \item \textbf{Are the findings from the mixed methods integrated?} One must ask  what are the similarities and divergencies in the findings (the output of the methods), asking does the integration add to more than the sum of the parts? Any contradictions or divergences should be explained or discussed. 
    \item \textbf{Are the different methods used rigorously conducted?} MMR studies should adhere to the rigor of the individual studies used in the mixed study (referring to the appropriate quality criteria for each study).\footnote{\edit{See the ACM SIGSOFT Empirical Standards for Software Engineering~\cite{ralph2020empirical}. Other references could be consulted for further considerations on the aspects of quality and validation in the MMR integration process, such as Teddlie and Tashakkori~\cite{teddlie_foundations_2009} and 
    Venkatesh \emph{et al.}~\cite{venkatesh2013bridging, venkatesh2016guidelines}.}} 
\end{itemize}

\subsection{Ethical Research} 
\label{sec:ethics}

\begin{center}
\noindent\fbox{
    \parbox{32.5em}{ 
\textbf{Principle of Ethical Research}: Mixed methods research studies should manifest the application of responsible research, with respect for people (e.g., participants, concerned parties, beneficiaries) and their welfare (e.g., respect for their confidentiality, privacy, time, emotions, cultural sensitivity, and needs), and that of the environment (e.g., organizations, communities of practice, society, and the planet at large, aiming for sustainability). By adhering to this principle, an MMR study report can address the question \textit{how responsibly did we do it?}
}}
\end{center}

Ethical research is a timeless concept, \edit{with a push by institutions and publishing organizations towards the use of Ethics Boards to review studies for ethical compliance.} However, the more recent unchecked advancements in AI and unprecedented access to publicly available datasets has meant that SE advancements are progressing faster than ethical standards, principles, and regulations can keep up. Both SE industry and research communities are calling for action to take stock of where we are headed collectively in terms of ethical practice in the software industry. Industrial and academic researchers alike need to be held accountable for, be aware of, and proactively manage the ethical aspects of research. 

An ethical approach to research starts by carefully considering the \textit{why} and not being driven by the \textit{why not}. This applies to the use of any method, but is even more pertinent to consider with mixed  methods because of the additional risks these designs bring in terms of privacy and confidentiality  when mixing and integrating different types of data.

\begin{itemize}
    \item \textbf{Considering the why:} Just because we may have access to large datasets and the latest in computational advancements, should we rush into designing research that can be performed using them or about them? What are the driving motivations for conducting the research in the first place? What benefits may it bring for the community being studied and for a wider society? Similarly, just because we can design a survey to ask some relevant questions, should we be spamming practitioner mailing lists with requests to fill them out? For example, contacting individuals, contributors, or leaders privately without explicit permission has been deemed as unacceptable behaviour by GitHub.\footnote{https://github.com/community/community/blob/main/CODE\_OF\_CONDUCT.md}
    
    \item \textbf{Privacy and confidentiality:} Deep and revealing conversations with research participants lend richness and nuance to the findings. Similarly, access to sensitive data for research purposes, such as team chats, emails, bug reports, and financial reports, is a privilege.  Typically, MMR studies collect and analyze a variety of data from multiple sources. While this is a strength of MMR studies, it is also a potential threat. Because of the multiple angles at which an MMR study approaches a phenomenon, mixing at the data level can place participant privacy and confidentiality under threat. For example, if recruiting survey participants for follow-up interviews from a survey cohort, it may be feasible to trace interviews back to survey responses using demographic data. Similarly, interview responses can be traced back to email, chat, or code commit logs, jeopardising the privacy and confidentiality of certain participants or data sources. What are our plans for protecting participant and data privacy and confidentiality when gathering, storing, and processing these assets, and when presenting findings based on them?

    \item \textbf{Respect and cultural sensitivity:} Many studies involve people from multiple locations and cultural backgrounds. Have we considered these cultural sensitivities when designing our protocols? Do we need to customise our protocols to make better sense to different participant groups or be more culturally appropriate in some communities?

    \item \textbf{Safety and welfare:} A cornerstone of ethical research is to ensure the safety and welfare of the people and environments touched by the research. For example, asking participants to recall adverse experiences or imagine challenging scenarios may cause stress. What are the physical, virtual, or psychological risks we may be exposing our participants and ourselves to? What are the environmental implications of our research? Do we have the requisite training to conduct risk assessments and to ensure the physical, psychological, and mental safety and wellbeing of our participants, ourselves, and the environment? 
    
\end{itemize}

\section{A Landscape of Mixed Methods Design}
\label{sec:landscape}

Research study design that involves choosing suitable methods and deciding on their timing is a challenging process~\cite{easterbrook_selecting_2008}, but research design is particularly challenging in the case of an MMR study. 
The research problem or questions being asked and how they emerge over the course of a study guides the research design and the timing of methods used.   
\edit{Below we discuss the main properties of research design, followed by different types of integration, and common patterns of MMR designs.} 

\subsection{\edit{Design Properties}}

\begin{itemize}
    \item \edit{\textbf{Research Questions:} 
    MMR questions differ from purely quantitative questions (that typically focus on description, comparison, or relationships between variables) or qualitative questions (that are typically open-ended, evolving, and exploratory with the aim to uncover processes or experiences). 
    The MMR question addresses what the researcher hopes to learn through the combination (integration) of the \qual~and \quan~data \cite{creswell_designing_2018}. In turn, these questions will shape other design properties as listed next.} 
    
    \item \textbf{Planned or Emergent:} 
    \edit{An MMR design can be either \textit{planned} (i.e., predetermined), where it is structured upfront based on the research problem or question, or \textit{emergent}, where it evolves organically throughout the study. The emergent approach may occur if the initial method proves ineffective (e.g., due to low participation) or if initial findings generate new insights that require confirmation, expansion, explanation, or triangulation.}

    \item \textbf{Inductive/Deductive Dominance:} An MMR design typically involves both \textit{inductive} (from data to theory) and \textit{deductive} research (from theory to data), but one approach usually dominates or takes priority over the other driven by the philosophical world view of the researcher or by the nature of the research problem or question. Researchers often conflate inductive and deductive research with \qual~and \quan~methods, respectively, but Young \emph{et al.}~\cite{young_spectrum_2020} explain 
    how inductive and deductive research may fall along a ``spectrum'' of \qual~and \quan~data. 

    \item 
    \edit{\textbf{Timing:} Time orientation in MMR design refers to the type of implementation process, specifically focusing on when \qual~and~\quan~data collection occurs. This can happen \textit{sequentially} (i.e., conducting first one \qual/\quan~method and then the other \qual/\quan) or in \textit{parallel} (i.e., conducting the two \qual/\quan~methods at the same time).
    For some cases, a sequential design is driven by the emergence of new or refined research questions, while a parallel approach may be motivated by the expertise of the research team, access to research participants and data, or timing constraints.} 
   
\end{itemize}

\edit{Additional design properties could be included in this list, such as \textit{epistemological perspectives} and \textit{paradigmatic assumptions} from a decision-making viewpoint, or strategies like \textit{sampling}, \textit{data collection} and \textit{data analysis} approaches (as also discussed in Venkatesh \emph{et al.}~\cite{venkatesh2016guidelines}). However, for the sake of simplicity, we focus on these key aspects: \textit{Research Questions}, \textit{Planned or Emergent}, \textit{Inductive/Deductive Dominance} and \textit{Timing}, complemented by a discussion on the \textit{Types of Integration} in the following subsection. Nevertheless, we acknowledge that all these design properties require careful consideration by researchers, as they significantly influence the choice of MMR design type (explained further in Subsection~\ref{sec:researchdesigns}).}

\subsection{\edit{Types of Integration}}
\label{sec:integrations}

Integration serves as a foundational element of MMR, as noted in Section~\ref{sec:background} and depicted in Figure~\ref{fig:interconnection},
and should be thoroughly detailed reporting any MMR study.
Creswell and Plano Clark endorse this view \cite[Chapter 2]{poth_sage_2023} and explain that a comprehensive description of integration, considering both its intent and procedural aspects, enhances transparency by demonstrating how different methods were effectively employed. Below we present four types of integration in MMR, as recommended by Kuckartz and R\"adiker \cite[Chapter 22]{poth_sage_2023}.

\begin{itemize}
   \item \textbf{Sequential integration} occurs when the findings from one method inform the application of another method. For instance, the results of an initial \quan~method may guide the development of \qual~data collection instruments or inform sampling strategies for the following phase. 
   
    \item \textbf{Results-based integration} involves combining, comparing, and relating findings only after the analyses of both \qual~and \quan~methods have been completed. This type of integration focuses on synthesizing the outcomes of each method (\qual~and \quan) to derive comprehensive insights.  
    
    \item \textbf{Data-based integration} involves the simultaneous analysis and interrelation of \qual~and \quan~data during the analysis phase. This type of integration requires that both data types are associated with the same underlying source, ensuring that \qual~and \quan~data can be distinctly assigned and connected to each individual case. This alignment allows for a more cohesive and in-depth examination of the research subject. 
    
    \item \textbf{Transformation-based integration} occurs when \qual~data is converted into \quan~data, or vice versa, allowing for integration within a single ``data world'' (\qual~or \quan). This type of integration enables researchers to analyze transformed data using the corresponding analytical techniques, enhancing the depth and breadth of the analysis. 
\end{itemize}

Additional examples of integration can be found in \edit{Onwuegbuzie and Johnson's Guide to Mixed Methods Analysis \cite{onwuegbuzie_routledge_2021}} and Poth's book \cite{poth_sage_2023}\edit{---the latter includes} a dedicated section on ``designing innovative integrations with technology'' organized by Guetterman, who introduced the idea of \textit{joint displays} during integration in MMR (see \cite{guetterman_integrating_2015, guetterman_visuals_2021}). Di Penta and Tamburri~\cite{Dipenta} also propose the adoption of \textit{mind maps} to gain an overview of the design to ensure there is alignment with the study's \quan/\qual~focus, and to aid in identifying gaps for further exploration, and to address threats to validity.
\edit{Venkatesh \emph{et al.}~\cite{venkatesh2016guidelines} provide more detailed prescriptive guidelines 
to assist in the integration and evaluation of MMR for Information Systems research, some of which may be useful for SE research.}

\subsection{Research Designs}
\label{sec:researchdesigns}

The \edit{design properties} mentioned above and how they are composed lead to a constellation of different research designs. 
We use the nomenclature for research designs as proposed by Creswell \edit{\emph{et al.}}~\cite{creswell_best_2011} while describing some popular design variations in the following. As we describe each design variation, we provide an example of a recent paper published in a relevant empirical software engineering journal that uses that design.\footnote{We found looking through the last twelve issues of the Empirical Software Engineering journal and the Information and Software Technology journal, that many SE papers use mixed methods, but they often do not say they are mixed methods research, and they also usually do not articulate or justify the specific design they use. Here, we provide one example paper that maps to each research design. If the paper was not that clear about the design, we  contacted the authors to confirm our mapping.}

\begin{itemize}
    \item  \textbf{Exploratory Sequential}: This sequential design often prioritizes a \qual~method (normally the dominant method) followed by a \quan~method that builds on insights from the first method. The design may be planned up front or it may emerge as insights are uncovered from the first method. 
    For example, Alami \emph{et al.}~\cite{Alami2024} explore how psychological safety impacts agile team behavior through a mixed method approach consisting of an initial exploratory phase with 20 interviews, followed by a \quan~survey study involving 423 participants. 
    
    \item \textbf{Explanatory Sequential:} This sequential design starts with and prioritizes a deductive \quan~method (normally the dominant method) that is followed by a \qual~method to help explain the insights from the first method. The design may be planned up front or the need for the second method may emerge, especially in light of some unexpected insights or where some insights need further interpretation or explanation. For example, Jiang \emph{et al.}~\cite{Jiang2024} mined open source deep learning projects to analyze the distribution of computer vision defects, followed by interviews conducted with leaders of a reengineering team, to shed light on the challenges and practices that may have led to the defects.     
    \item \textbf{Convergent Parallel:} The convergent parallel design involves both a \qual~method and a \quan~method to bring complementary insights on a research question. The goal may be to uncover both convergent and divergent insights from both methods that are planned in parallel. For example, Hidellaarachchi \emph{et al.}~\cite{Hidellaarachchi2024} carry out personality test-based surveys and in-depth interviews with a selected set of participants to investigate the impact of personality on requirements engineering activities. The survey and interviews were conducted in parallel and insights obtained from both approaches were used to answer the research questions. 
    
    \item \textbf{Embedded:} \edit{An embedded design nests a secondary method within a primary method with the intent up front to integrate the data from both methods to arrive at the study findings (that is, it relies on data-based integration as opposed to results-based integration we may see in a sequential design). 
    In the study by Madugalla \emph{et al.}~\cite{Madugalla2024} which 
    aimed to identify challenges and possible benefits of conducting SE research with human participants during the pandemic, research findings emerged from the integration of the data obtained with a survey (\quan~and dominant method) with data from the interviews conducted with a sample of the same participants (\qual~and nested method). Yu and Khazanchi~\cite{yu2017} offer practical advice for conducting embedded MMR designs. 
    }

\end{itemize}

\section{Mixed Method Research Design Scenarios}
\label{sec:scenarios}

To illustrate the principles, properties and MMR designs we presented so far in this \manuscript, we showcase four scenarios as guiding examples. There could be many more study design examples, but we focus on four that map to each of four MMR designs (exploratory sequential, explanatory sequential, convergent parallel, and embedded). For each scenario, we suggest a fictional SE researcher who is pursuing a mixed method design, and use these characters to show how having empathy and understanding for our study participants can also lead to more thoughtful and rigorous research. 
We describe the four scenarios below and mention the decisions and challenges each fictional researcher could have faced in designing and conducting their research in terms of the four principles we summarize in the paper. For the four scenarios, we provide illustrations of the research designs (see Figures \ref{fig:ali}, \ref {fig:sam}, \ref{fig:vicki} and \ref{fig:zara}). We also summarize the four scenarios in Table \ref{tab:personas}. \edit{The study scenarios do not mention all the necessary details expected in a paper, as we focus on illustrating the principles and properties of mixed method research in SE to match the goal of this paper.}  

\begin{table}[h]
\resizebox{\columnwidth}{!}{%
\begin{tabular}{lllll}
\hline

\textbf{Scenario} & \textbf{Type of Design} & \textbf{Design Flow} & \textbf{Rationale(s)} & \textbf{\edit{Integration}} 
\\ \hline

\begin{tabular}[c]{@{}l@{}}\ref{sec:sce_ali} \textbf{Ali} \\ Security Compliance\\ Tool\end{tabular} & \begin{tabular}[c]{@{}l@{}}Exploratory Sequential \\(planned)\end{tabular} & \begin{tabular}[c]{@{}l@{}}quaL: Inductive Interviews $\rightarrow$ \\ quaN:  Deductive Survey\end{tabular} & \begin{tabular}[c]{@{}l@{}}Expansion\\ Triangulation\\ Development\end{tabular} & \begin{tabular}[c]{@{}l@{}}\edit{Sequential} \end{tabular}
\\ \hline

\begin{tabular}[c]{@{}l@{}}\ref{sec:sce_sam} \textbf{Sam} \\ AI Code review\end{tabular} & \begin{tabular}[c]{@{}l@{}}Explanatory Sequential \\(emergent)\end{tabular} & \begin{tabular}[c]{@{}l@{}}quaN: Deductive Data Mining $\rightarrow$\\ quaL: Inductive Interviews\end{tabular} & \begin{tabular}[c]{@{}l@{}}Explanation\end{tabular} & \begin{tabular}[c]{@{}l@{}}\edit{Sequential} \end{tabular}
\\ \hline

\begin{tabular}[c]{@{}l@{}}\ref{sec:sce_vicki} \textbf{Vicki} \\ Adoption of \\ Onboarding Mentorship \end{tabular} & \begin{tabular}[c]{@{}l@{}}Convergent Parallel \\(planned)\end{tabular} & \begin{tabular}[c]{@{}l@{}}quaL: Inductive Observations $|$ \\ quaN:  Inductive Data Mining\end{tabular} & \begin{tabular}[c]{@{}l@{}}Complementary\end{tabular} & \begin{tabular}[c]{@{}l@{}}\edit{Results-based} \end{tabular}
\\ \hline

\begin{tabular}[c]{@{}l@{}}\ref{sec:sce_zara} \textbf{Zara} \\ Test generation \end{tabular} & \begin{tabular}[c]{@{}l@{}} Embedded \\(planned)\end{tabular} & \begin{tabular}[c]{@{}l@{}}quaN: Deductive Lab Experiment $|$\\ quaL: Deductive Surveys\end{tabular} & \begin{tabular}[c]{@{}l@{}}Expansion\\ Credibility\end{tabular} & \begin{tabular}[c]{@{}l@{}}\edit{Data-based} \end{tabular}
\\ \hline

\end{tabular}%
}
\caption{Four fictional research scenarios demonstrating four different possible designs for mixed method research\edit{, showing the flow of the methods, the rationale(s) and type of integration for each design.}} 
\label{tab:personas}
\end{table}

\subsection{Ali's Study: Investigating the impact of a new security tool on developer productivity and developer experience at a large software company.}
\label{sec:sce_ali}

\color{black}

\textbf{Ali} (he, him), is a postdoc working closely with a partner company studying the introduction of a new security tool to ensure compliance with secure development practices across a large software company. 
Ali and his industry partner recognize that introducing a tool that enforces changes in development work has complex and likely unexpected impacts, not just on the security and or quality of the authored code, but also on the developers' perception of their productivity and experience.  His \textbf{research goal} is to identify what impacts the tool may have (within the context of the developers' work and environment), and using these initial insights, investigate if these perceptions are shared broadly by other developers across the company. 
Ali's \textbf{research question} asks: How does a development tool that enforces compliance with secure development practices impact developer perceived productivity and their experience?

\begin{figure}[th]
    \centering
    \includegraphics[scale=0.18]{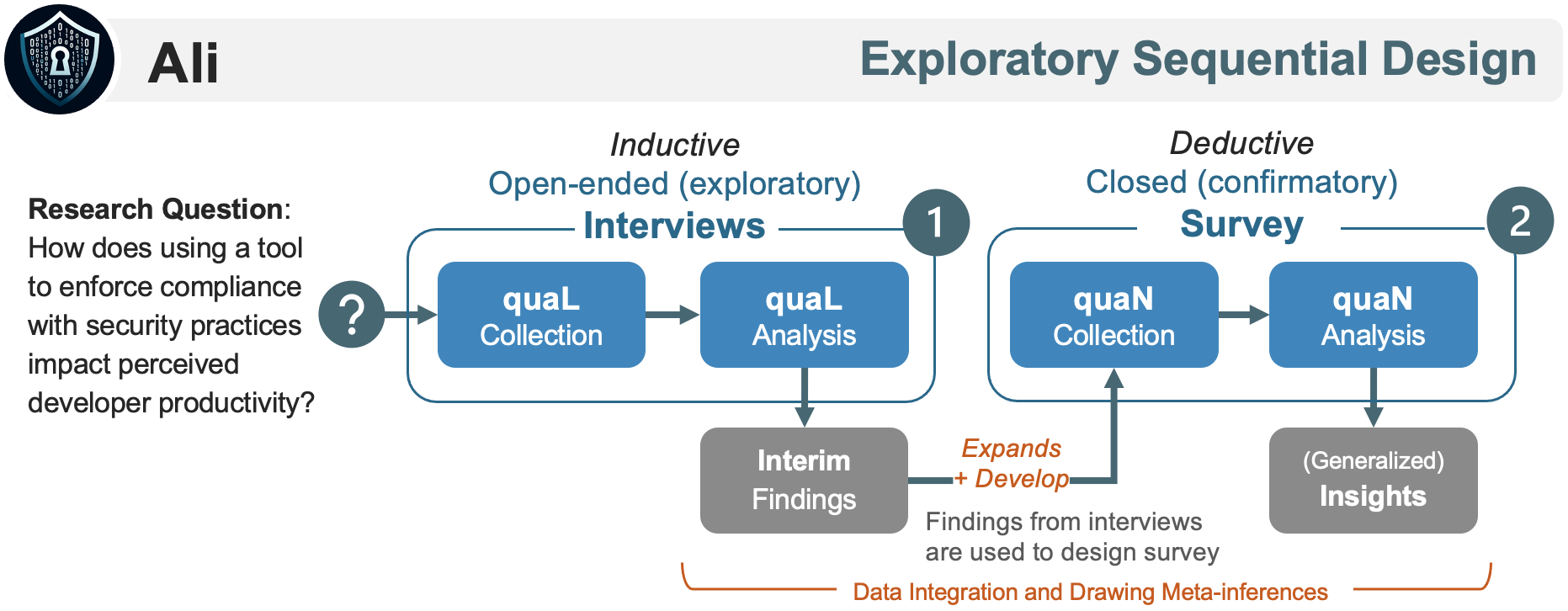}
    \caption{Ali follows an exploratory sequential design, studying the impact of a security tool that enforces secure development practices. The unit of analysis are the practices informed by the developers using the tool across a company.}
    \label{fig:ali}
\end{figure}

\paragraph{Methodological Rationale:}
Through a literature review, Ali finds a lack of research on developers' perceived impacts of the specific type of security tool he is studying. He aims to investigate whether developers believe that the tool improves the quality of the code but also how it may impact other aspects of their development experience. 
Ali could begin by designing a mono-method survey, such as one based on a validated technology adoption model~\cite{tam89}, and distributing it widely across the company. However, his primary goal is to explore \textbf{inductively} \emph{what impacts} the new security compliance tool might have, focusing on both the anticipated and unanticipated aspects of developers' experiences. Furthermore, Ali aims to assess whether these initial findings can be \textbf{generalized} to a broader population of developers within the company. Ali's goals could not be achieved without an integrative thinking approach supported by MMR, which leads him to use an \textbf{exploratory sequential} mixed methods design.
First, Ali interviews (\textbf{\qual~}method) about 20 developers to uncover these impacts. Next, based on the initial insights, he conducts a follow-up survey (\textbf{\quan~}method) across a broader population of developers at the studied company.
A survey that is faster to answer was part of his rationale for this design. See Figure \ref{fig:ali} for an overview of Ali's MMR approach.

\paragraph{Novelty of Integrated Insights:}
The \textbf{sequential integration} of findings from both methods provided novel insights into the impact of a new security compliance tool on developers' perceived productivity, workflow, and learning experiences. Furthermore, these impacts were found to be generalizable to other developers.
Through interviews, Ali finds the tool does not negatively impact perceived developer productivity in terms of velocity, and that it surprisingly increases the developer's flow experience and helps the developers learn about and appreciate new security practices. He also uncovers that some developers abandoned using tools they were fomerly using because of this new tool, another finding he did not expect.
These anticipated and unanticipated findings were used by Ali to \textbf{develop} his follow-up survey with closed survey questions to investigate quantitatively if these experiences generalized across a broader set of developers at his partner company. 

\paragraph{Procedural Rigor:}
The interviews and survey were conducted  following respective best practices.\footnote{We do not elaborate on the rigor of the individual methods used to focus our paper on the discussion of rigor that applies to the mixing of methods.} As mentioned above, the rationale for using these two methods was primarily to \textbf{develop} the survey and to \textbf{triangulate and generalize} the findings across a broader sample. Ali aimed to conduct both methods (sequentially) at the start of the research, but considers the qualitative insights as the \textbf{dominant} contribution in this \textbf{exploratory sequential} research design.   
Ali was careful not to invite those interviewed to answer the survey, as doing so could influence their survey responses. Ali also took care to word survey questions to avoid bias with insights gained through the interviews. 
Ali assesses the overall findings from MMR, not from individual studies but as an integrated contribution. He discusses the potential threats to validity that may arise during data collection and analysis when compiling the final results report for his study. For instance, 
it could have been the case that the follow up survey led to divergent results about the abandonment of other tools or on productivity experiences. Such divergences would need further investigation.   

\paragraph{Ethical Considerations:}
As Ali conducts both methods, he takes care to protect the identity of the team members he interviewed and surveyed. Ensuring this confidentiality was especially important given security is a key concern of the organization, and developers may not feel free to share concerns about the friction a compliance tool may introduce. Ali was also careful how nuanced insights from the interviews (e.g., about abandoned tools which may have been used by specific developers only) were later framed in a neutral ways so that the small group of interviewed developers could not be identified.  Ali also took care not to send the survey to the interviewed participants, both to ensure the integrity of the data (supporting rigor), but also to be respectful of  participant time.  The design itself was also done with the careful use of developer time in mind.    
\color{black}

\subsection{Sam's Study: Studying The Impact of a Generative AI Tool that Automates Coding Reviewing}
\label{sec:sce_sam}

\color{black}

\textbf{Sam} (they, them) is an early-stage Ph.D. student collaborating with a large company that is trialing a new Generative AI tool to automate part of the code review process.  
Sam's initial \textbf{research goal} is to test their \textbf{hypothesis} that a new Generative AI (GenAI) tool that automates the code review process will improve the time to merge of code changes and improve the quality of the reviewed code (by reducing the number of bugs in the newly committed code). 

\begin{figure}[th]
    \centering
    \includegraphics[scale=0.125]{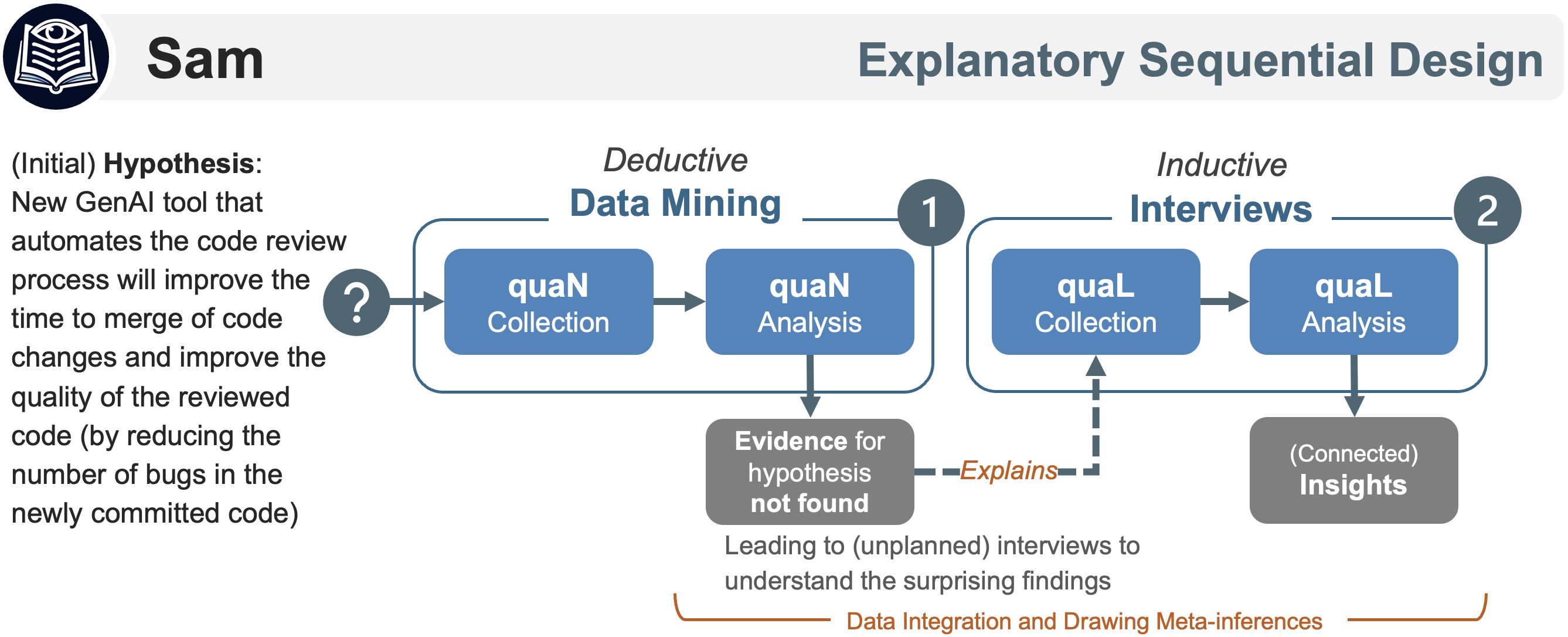}
    \caption{Sam follows an explanatory sequential design to study the adoption of a generative AI reviewing tool on code quality.  The unit of analysis is telemetry process metrics in (1), while in (2) it is perception of the process informed by developers. }
    \label{fig:sam}
\end{figure}

\paragraph{Methodological Rationale:}
Sam starts their research with a \textbf{deductive} data mining method (\textbf{\quan~}method) to investigate their hypothesis that the GenAI reviewing tool (before and after use) improves time-to-merge and reduces the number of bugs captured in their industry partner's code repository. They do not start with the intention to follow a mixed methods research design, and were planning to conduct the data mining study only. 
However, from the mined data Sam confirms the tool leads to the hypothesized faster time-to-merge but not a lower number of bugs after merge. Sam wishes to understand this unexpected result.  To gather the views of the developers who used this tool, Sam feels compelled to conduct open ended interviews (\textbf{\qual~}method) \textbf{sequentially sampling} a subset of the developers to probe \textbf{inductively} why the tool did not have the hypothesized impact on bug count. The resulting research design Sam followed is an \textbf{explanatory sequential design}  (see Figure~\ref{fig:sam}).

\paragraph{Novelty of Integrated Insights:}
From the interviews combined with the mined data, novel insights about the tool's efficacy and how it was used, emerged.  Notably, from the \textbf{sequential integration} of the mined data and interviews, Sam learns that the developers are able to work more efficiently but they are potentially more complacent when using the GenAI reviewing tool.
Sam reflects on the mixed insights and their implications, limitations, and future directions.
For instance, these integrated insights can be used to shape code review tool design accordingly, by adding features to prompt developers to avoid complacency. 

\paragraph{Procedural Rigor:}
A mixed methods design was not intended up front, but was an \textbf{emergent} or organic research design decision to \textbf{explain} the surprising results from the data mining. 
In addition to ensure the mining process and interviews were conducted with rigor, Sam must ensure that the \textbf{sequential integration} step was also done in a way to reliably explain the surprising result from the mined data. As Sam may be biased following this unexpected finding, they recruit another student to conduct these interviews.
Note Sam could have used an \textbf{embedded} design with a minor survey strand to explain the insights from the mined data but the findings were so unexpected they anticipated that interviews could also inform how the review tool could be improved before more mining.  

\paragraph{Ethical Considerations:}
As Sam mines the telemetry data, they ensure that their findings do not reveal whose coding or code review activities may have resulted in the future introduction of bugs with any other team members or managers, as this could be used to evaluate their performance.   
Care is also taken by Sam to protect the identity of the developers interviewed, and also to ensure the interviews did not make participants feel stressed given the presence of some bugs they may have introduced that was found in the telemetry data and that they feel could be linked to them directly. 

\color{black}

\subsection{Vicki's Study: Understanding the Adoption of a New Onboarding Process in a Large-Scale Open Source Project}
\label{sec:sce_vicki}

\color{black}

\textbf{Vicki} (she, her) is an experienced researcher that has been studying collaboration practices in open source projects. Her most recent \textbf{research goal} is to understand the adoption of a new onboarding process to address the retention of new contributors to an open source project that she has been involved with (as a contributor and researcher) for several months. 

\begin{figure}[th]
    \centering
    \includegraphics[scale=0.18]{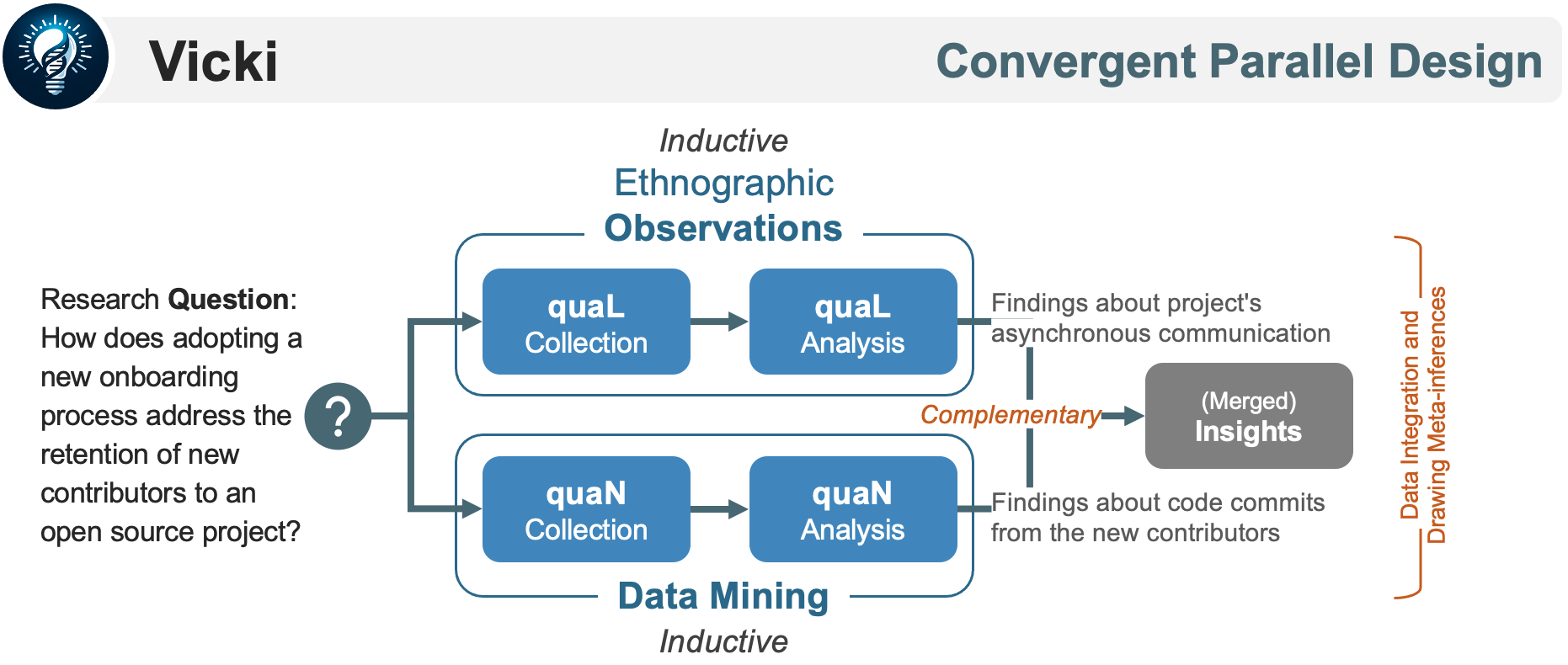}
    \caption{Vicki follows a planned convergent parallel design to study the adoption of a new onboarding process in an open source project.} 
    \label{fig:vicki}
\end{figure}

\paragraph{Methodological Rationale:}
The onboarding process relies on a style of mentoring that has not been studied before and suggests that seasoned and new developers loosely pair on work items, asynchronously communicating over Slack. 
As this is a new process, Vicki decides to \textbf{inductively} consider the onboarding process from two angles in \textbf{parallel}. Vicki uses observations (\textbf{\qual~}method) of participant communication over the project's asynchronous communication channel to discern how the onboarding process of pairing is used by both existing and new contributors in the project.
In parallel to, and to \textbf{complement} her observations, she mines logged data (\textbf{\quan~}method) about code commits from the contributions that are made by the new contributors. She studies the frequency and timing of commits during and following the informal pairing activities over Slack. 
See Figure \ref{fig:vicki} for an overview of her MMR design. 

\paragraph{Novelty of Integrated Insights:}
Vicki first analyzes the qualitative and quantitative data to understand the nature of the communication and the impact of onboarding on commits. She then conducts a \textbf{results-based integration} by combining, comparing and relating findings from the qualitative communication data and the quantitative commit data. 
This integration leads to the meta-inferences that contribute theoretical insights about how pairing during the onboarding process may lead to faster commits.
The use of a \textbf{convergent parallel design} that involves conducting both \qual~and \quan~methods in parallel provides complementary and \textbf{novel insights} on the adoption of this new onboarding process in this open source community. Vicki also finds that the onboarding mentorship process of pairing and communicating over Slack leads to an increase in future contributions and higher quality commits.  

\paragraph{Procedural Rigor:}
 This design allows Vicki to integrate and build a story about how the mentors and new contributors communicate and pair, and the impact of that communication on the timing and presence of subsequent or in parallel commits. She considers the quantitative aspects (number and timing) of the commit data in an inductive way to see what emerges (she has no initial hypothesis about the impact of this informal pairing over Slack). As she integrates the two data sets, she is careful to specify the timing intervals as she links the two sets of data (the timing interval used is suggested by other research that links these two types of data).  Together, these two data sources lead to complementary insights and theoretical insights about the nature of the communication about pairing and the commits the communication is connected to. Her analysis leads to meta-inferences about the pairing activities and commit behaviour, notably that pairing during onboarding lead to faster commits. The rigor of her design is increased by checking the coding reliability with other researchers and running statistical analysis on the quantitative data she collects.

\paragraph{Ethical Considerations:} 
Vicki seeks permission from her research participants and project owners to both monitor the project communication channels and mine logged data about code commits and their comments. She also ensures that new contributors agree to having their communications with the mentors analyzed for the study, and that they may opt out at any time (ensuring any collected information is deleted from the study data across both methods). Likewise, mentors also must give permission and can opt out at any time. 
She anonymizes the identity of the participants she observes in the communication channel, ensuring no identifying data remains in the reported data. 
%
%
Vicki recognizes that mining the commit data comes with the potential risk of introducing significant privacy concerns, so she also obfuscates the commit data. 
Vicki conducts her observations of the Slack channels in a \textbf{transparent but nonobtrusive} manner, ensuring security in its storage and planned deletion after the study.

\color{black}

\subsection{Zara's Study: Evaluating the Effectiveness of an Automated tool for Generating Tests}
\label{sec:sce_zara}

\color{black}

\textbf{Zara} (she, her) is a Ph.D. student who aims to investigate how effective an automated tool she designed is for generating tests.  Zara faces the challenge of evaluating a tool prototype she developed without an industry partner, and she does not have access to developers in industry for evaluating her tool.   Her \textbf{research hypothesis} is that her tool will lead to faster test generation when used by senior computer science students. 

\begin{figure}[th]
    \centering
    \includegraphics[scale=0.18]{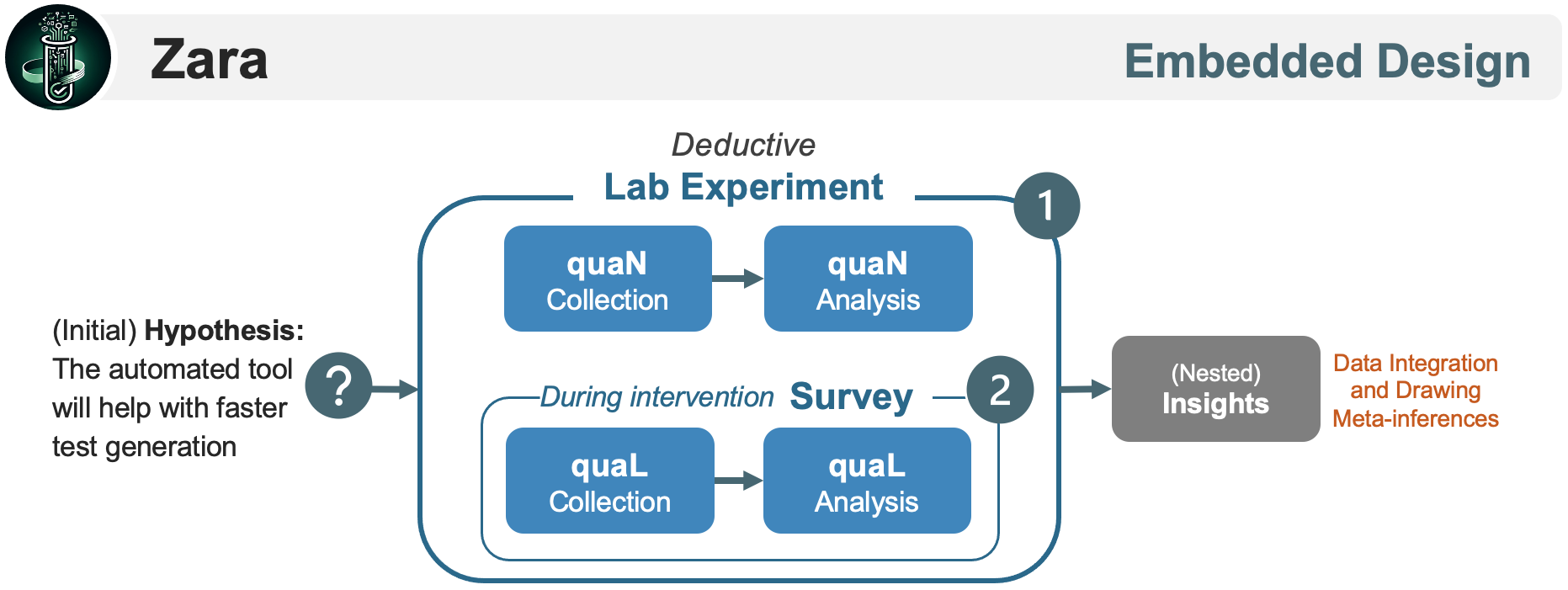}
    \caption{Zara follows a mixed methods embedded design to study a test generation tool. The unit of analysis is the test cases generated by the tool, in addition to the experience of the participant using (or not using) the automated tool. }
    \label{fig:zara}
\end{figure}

\paragraph{Methodological Rationale:}
Zara starts, deductively, with the hypothesis that use of her tool will lead to faster test generation over standard test generation features available in the IDE. Zara decides to conduct a lab experiment (\textbf{\quan~}method) with senior student developers who are well versed in software testing techniques, where half of the students she recruits use the tool and the other half do not. 
Zara also anticipates that the students may have more confidence about the tests they authored and may learn about testing through use of her tool. Inviting all participants, she decides to run an embedded data collection method, where she asks them to describe their confidence about the tests they create, and how the tool or IDE they are using impacts their knowledge about testing through intermittent short surveys (\textbf{\qual~}method) presented between tasks.   Without this embedded qualitative method, she would not gain insights into their confidence of their tests or how they learned more about testing.
The \textbf{embedded deductive design} she used is primarily a quantitative method, examining the speed of tests created, but has a smaller embedded deductive qualitative component, to report on their perceived higher levels of confidence and learning. Note that the secondary method relies on the dominant method leading to the embedded design decision. See Figure \ref{fig:zara} for an overview of this MMR design. 

\paragraph{Novelty of Integrated Insights:}
Zara's study reveals that using the automated tool for test generation not only improves the speed of test creation compared to standard IDE features (identified from the experiment) but through the embedded survey she finds it also enhances the participants' confidence in the tests they generate and their overall understanding of software testing concepts.
Using a \textbf{data-based integration} that involves the simultaneous analysis of both quantitative and qualitative data brings this new meta-inference about the impact of this new tool (when compared to the features in an IDE) on student perceived confidence and learning. 

\paragraph{Procedural Rigor:}
As Zara designed the study, Zara invited two other graduate students in her lab to validate the data collection instrument developed/used. 
Participants are selected randomly for each condition, and those who are not assigned to use her tool are free to use other test generation features available in modern IDEs. Zara assigns the same testing tasks to both groups in this between groups design, and asks similar questions to both groups in the embedded survey to tease out how the used tools impacted their confidence in testing and their learning of testing concepts. 
The advantage of using the embedded survey, is that it helped explain, and lead \textbf{credibility} to the quantitative results from the mined data.

\paragraph{Ethical Considerations:} 
The use of student participants calls for several ethical considerations.  Participation is voluntary and their identity is kept confidential across both methods. Zara takes care during recruitment that the students' teachers are not involved in the study or research, that the teachers do not know who participates in the study, and that they do not put any real or apparent pressure on  students to participate and thus do not introduce power dynamics. The students are reassured that their decision to participate (or not) and nature of the feedback they provide will have no bearing on their performance in the class.
The students' time to participate is also valued. The potential benefit of gaining experience with test generation is communicated to the students during recruitment (this value was validated with teachers before the study).  Care is taken to ensure students do not feel stressed during the study, and they can opt out at any time. 


\color{black}

\subsection{Learning from Other Examples}

So far, we have shared that using MMR enables SE researchers to study their research topics from a more holistic perspective, expanding and complementing methods to achieve more novel insights. We also note from the scenarios we provide (and the paper examples we reference in Section \ref{sec:researchdesigns}) that mixed methods are particularly valuable in studying socio-technical aspects of SE phenomena. The research scenarios we provided help to demonstrate the broad and varied landscape of possible MMR designs in SE research. The ACM SIGSOFT Empirical Standards for Multiple and Mixed Methods also provides a list of MMR papers~\cite{ralph2020empirical}, but care has to be taken in using these and other examples of MMR designs as ``exemplars'' to guide future work. With this in mind, we next present ``antipatterns'' to avoid or to watch out for when designing or critiquing an MMR design. An earlier version of these antipatterns also appears in the ACM SIGSOFT Empirical Standards, contributed by the first author of this paper.

\section{Antipatterns of Mixed Method Research Designs}
\label{sec:antipatterns}
 
The principles and research design patterns we described above can be followed when designing a study.  And through the scenarios, we illustrated how the principles of MMR could be followed.  In the following, we provide examples of ``what not to do'' and ``what to avoid''. 

As researchers, authors and reviewers of papers, we have noted a number of common antipatterns in how mixed methods are used or presented.
Some of the antipatterns we encountered originate in issues with how the research is presented, rather than inherent to the conduct of the research. We introduce \textit{presentation} antipatterns first noting how they can be potentially fixed.  The other \textit{study design} antipatterns are harder to ``fix'' after the research has been conducted, but hopefully can be identified at design time (or during review of a paper).  We recommend researchers consider these antipatterns early in their research process.

\subsection{Presentation Antipatterns}

As mentioned, these antipatterns are concerns in how the research is presented or written.  

\begin{itemize}
 \item \textbf{Uninvited guest or party crasher:} This antipattern occurs when one of the methods used is not clearly introduced in a study report introduction or methodology and makes an unexpected entrance in the later part of the report, such as in the discussion or limitation sections. To fix this, the author should clearly mention all methods in the introduction and methodology sections of the study report, justifying why they were used. 
 Even with emergent designs, as in the case of Sam~(Sec.~\ref{sec:sce_sam}), all methods used should be presented early in the paper with a mention that the design was emergent.  

\item \textbf{Smoke and mirrors:} This antipattern can be spotted when a paper attempts to oversell a study as an MMR when one of the methods merely offers a token or anecdotal contribution to the research motivation or findings. This can be fixed by clarifying the presentation in the paper. For example, in the case of the embedded design carried out by Zara~(Sec.~\ref{sec:sce_zara}), if the treatment of the embedded part of the study (to find out how the tool impacted confidence in testing) was shallow and the number of participants surveyed was very low, the paper would have oversold the contribution as an MMR study.

\item \textbf{Limitation shirker:} This antipattern occurs when there is a failure to discuss limitations from all methods used or from their integration in a study report. MMR will have a much longer and more in depth limitation section than a single study paper, and the limitations reported should cover all methods used and be congruent with the methods mixed, as well as address how the data or analysis was integrated. In the case of Vicki~(Sec.~\ref{sec:sce_vicki}), failure to describe the limitations in how participants were observed (in this case, perhaps other forms of communication occurred outside Slack) would weaken the conclusions from integrating the insights from the observations of the commit behaviour.  

\end{itemize}

\subsection{Study Design Antipatterns}

These antipatterns are ingrained in how the study was designed or conducted. They are hard, if not impossible to fix afterwards, but thinking about them early on can help researchers avoid them.

\begin{itemize}

\item \textbf{Missing the mark:} This antipattern occurs when the MMR design is not aligned with the research question or objective. This can be seen when one or more of the methods do not make sense in light of the research question being posed.
The fix for this should happen at the study design phase by asking if every method answers the core research questions being asked. \edit{ In the case of Zara, the research question considered not just the testing speed, but also testing confidence. If the embedded method did not lead to insights about testing confidence through the data integration step, this design could have been described as ``missing the mark''.  }  

\item \textbf{Selling your soul:}  This antipattern is related to ``missing the mark'' and may occur when a researcher  employs an additional method to appeal to a future reviewer who they believe has preference for a particular methodology or study outcome, but the use of the method does not contribute substantively to the research findings. This can be addressed by using the design one authentically believes in, or by ensuring that the researcher draws out meta-inferences from the different methods used. In the case of Ali~(Sec.~\ref{sec:sce_ali}), the findings from the interviews may have been sufficient to stand on their own merit.  But Ali may also have been influenced to do a \quan~survey only to appeal to a reviewer that may be focused on knowing if the findings generalized to more teams in the company.

\item \textbf{Cargo cult research\footnote{Cargo cult can be a metaphor for superficial imitation of a process without basic understanding of its mechanism.}:} Using research methods in which the research team lacks expertise but that they have observed being used to generate ``exciting findings'' in well cited papers that use these methods. Collaborating with other researchers with broader expertise may help avoid this antipattern.  For example, in the case of Vicki, adding a data mining method to the observation part of the design may require working with collaborators with mining expertise if Vicki lacks that expertise.

\item \textbf{Sample contamination:} This antipattern may occur in an MMR sequential design where the same participants are used in multiple, sequential methods without accounting for potential contamination from earlier method(s) to later ones (where participants have already seen an intervention or know about the research). For example, in the case of Ali, Ali should be careful not to survey the participants already interviewed.
However, in some cases the design itself may involve selecting participants (e.g., for interviews) from an earlier method (e.g., a survey) and the resulting ``contamination'' is intentional (for example, in the case of an embedded design). Any possible  contamination should be justified or reported, and mentioned in the limitations of the study report. 

\item \textbf{Lost opportunity:} This may occur in an MMR sequential design when there is a failure to use findings from an earlier method during the development of an instrument for a later method. This should be caught during the implementation of the study as it cannot be fixed through the presentation (alternatively, the studies can be written up as a two-phase study or as two separate studies with two different contributions).  In the case of Ali, not using insights from the interviews to ask pertinent questions in the survey about the security tool mandate could be a lost opportunity.

\item \textbf{Integration failure:} We talked about failure to integrate data or insights as a quality attribute earlier in the \manuscript. This antipattern occurs when there is poor integration of findings from all methods used in the study design. This can be fixed by planning how the results/findings will be integrated and making sure that the results/findings from both methods are integrated or merged in the results/findings section of the paper (or earlier in the case of an embedded design). In the case of Zara's research, it may be difficult to see how to integrate the insights from the experiment and embedded survey. The insights of how the students became faster over time in the experiment, and were learning as gleaned from their self reports, can be integrated to emphasize the insight about learning using the automated testing tool and add credibility to the insights about effective testing from the dominant study. It may also be the case that the researcher integrates the findings from the mixed methods, but fails to adequately report them in the presentation (note, we could call this antipattern ``integration presentation failure'').

\item \textbf{Questionable ethics:} Finally, even when all of the above antipatterns are avoided, i.e., the MMR study is applied and presented well, it may come across as having questionable aims, or as using questionable means to achieve the aims.  The researcher should clearly state the motivation for the study and expected benefit to society or research to help justify the possible risks a study may involve. In the cases of all the fictional researchers above, we mentioned at least some of the ethical issues to be considered. We refer the reader to the points raised about conducting MMR ethically in Section~\ref{sec:ethics} and encourage all studies should undergo an ethical review.
 
\end{itemize}

\section{Discussion}
\label{sec:discussion}

In this section, we discuss how MMR relates to other methods (such as multi-methods) and present a call for action.

\subsection{Synergies with Other Methods}
\label{synergies}

Mixed methods research can easily be confused with multi-methods research.\footnote{We remind the reader this and other terms are defined in the Glossary found in the Appendix} The difference between mixed methods and multi-methods lies in \textit{what} types of data are used and \textit{how} the data is used. Mixed methods requires the integration of quantitative (\quan) and qualitative (\qual) data. Multi-methods can be applied with any combination of \quan-\quan~or \qual-\qual~methods. The principles we outline in Section \ref{sec:principles} all apply to multi-method research as well.

Mixed methods are often used within or in combination with other research methods. For example, case studies \cite{runeson2012case},  action research \cite{Staron20191}, socio-technical grounded theory \cite{hoda_socio-technical_2022}, ethnography \cite{sharp2016role}, or experiment based studies \cite{wohlin2012experimentation} can be conducted with a mixed methods research approach.

Finally, although the focus of this paper remains on primary studies, secondary research, such as systematic reviews of the literature (SLRs) \cite{Kitchenham2023SEGRESS, PagePRISMA2020}, families of experiments \cite{Basili1999456,Santos2020566}, and rapid reviews \cite{Cartaxo2020, Bjarnason2023} that ground their application to methodological aspects and evidence synthesis using a mixed methods approach, can also benefit from the principles we synthesize in this paper. %

\subsection{A Call to Action}

\begin{figure}[th]
    \centering
    \includegraphics[scale=0.5]{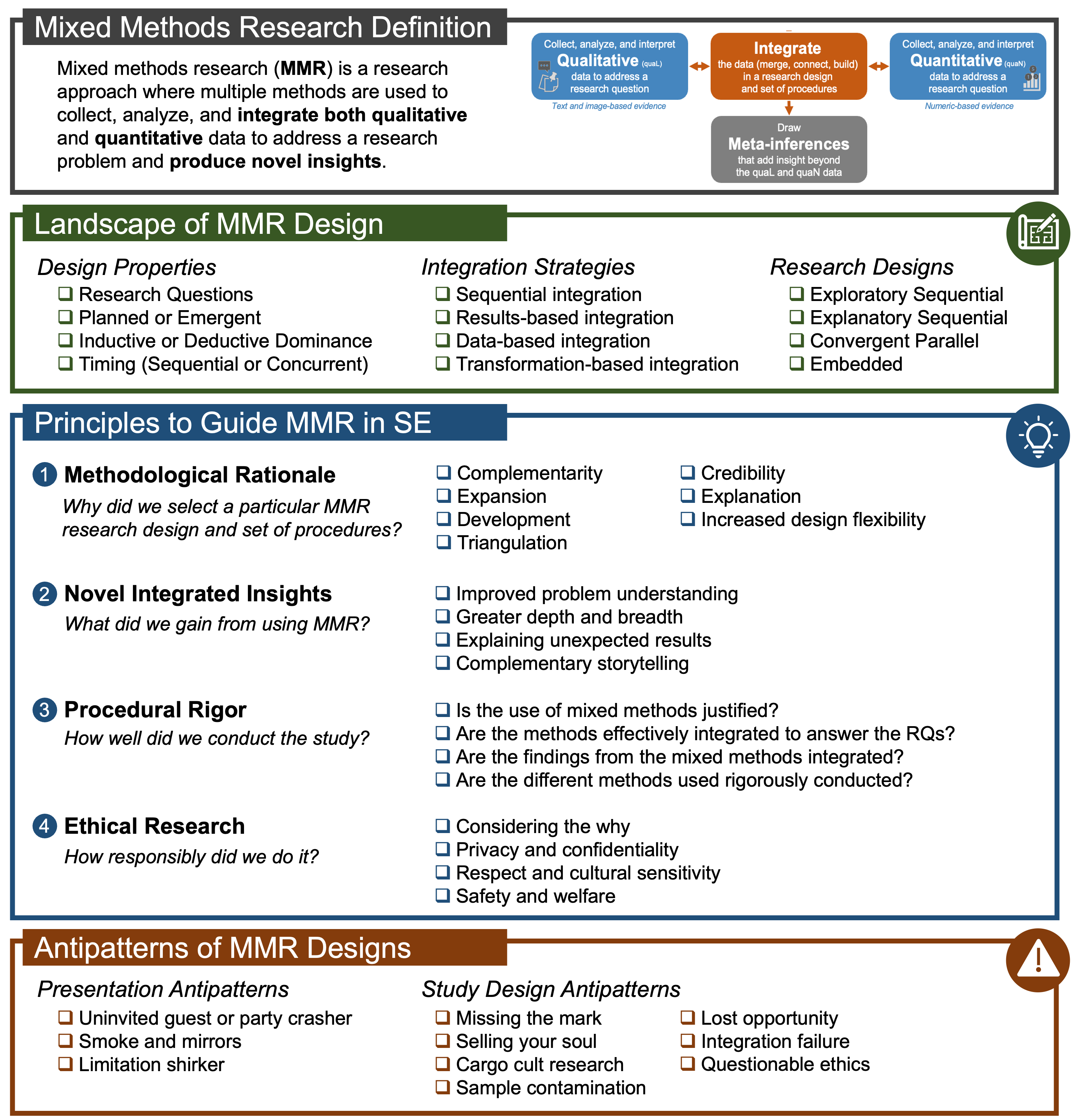}
    \caption{\edit{Mixed methods research in SE: definition, design landscape, principles and antipatterns.}} 
    \label{fig:summary}
\end{figure}

\edit{
Much of SE research is inherently socio-technical and as such can benefit from the use of MMR to bring insights on the underlying context and to strengthen the theoretical contributions from our work, and capability to generalize or translate our findings to other contexts.  We call for a principled use of mixed methods in SE research.  
Figure~\ref{fig:summary} summarizes the key points we raised in our paper: the MMR definition, a landscape of MMR designs, four principles to guide MMR in SE, and a list of antipatterns of MMR designs to be considered.} 

In adapting the \edit{guiding principles} we propose for an SE context, we do not simply promote MMR but \textbf{emphasize using MMR ethically}, as captured in the \textit{principle of ethical research}. As a research community we have conducted large-scale surveys with developers and mined developer forums which in some cases has led to outrage in the developer community and introduction of new policies for unsolicited research calls for participation (see Section~\ref{sec:ethics}). As researchers, we should have empathy for the people we are studying, and indeed aim to help them. They are real people. This is another reason why we include examples of fictional researchers, painting a picture of these researchers as human beings and the struggles they may face. Many of these struggles are not shared in the articles they write but perhaps could be so we can learn from each other. We recommend using MMR to make SE research more human focused--amplifying the human side of the quantitative data with the inclusion of qualitative data, both collected ethically.

As the lure of AI tools tempts the SE research community with promises of easing or accelerating various research tasks and methods, it is timely that we get into the habit of \textbf{asking ourselves `why', and not just be driven by the temptation of `why not'}, as captured in our \textit{principle of methodological rationale}. At the same time, as we begin to explore the risk and the opportunities of using LLMs for qualitative research~\cite{bano2023LLM, hoda2024Book}, let us \textbf{remind ourselves of the value of `deep work'}, captured in the principle of \textit{procedural rigor}. We strongly believe that using MMR studies conducted in interdisciplinary research settings can help address complex, socio-technical research questions and \textbf{drive meaningful research}, captured in our \textit{principle of novel integrated insights}.

These \edit{guiding principles} for using MMR in SE are required now more than ever. We hope this first step will help present and future SE researchers to use mixed methods research ethically, rationally, and rigorously to generate novel insights, \edit{and encourage our community to develop more prescriptive guidelines and playbooks for using and evaluating mixed methods research in SE.}

\section{Conclusions}
\label{sec:conclusions}

In this paper, we provide an overview of how mixed methods research designs can be used in software engineering, with the intention to inspire their use.  
Other resources, such as the ACM SIGSOFT Empirical Standards~\cite{ralph2020empirical}, as well as an earlier paper by Easterbrook \emph{et al.} 
on selecting a research method in software engineering~\cite{easterbrook_selecting_2008}, only lightly touch on the concepts and challenges of using mixed methods research in software engineering. As can be seen from Section \ref{sec:background}, there is a long history of research behind the mixed method paradigm and \edit{use in other disciplines, some related to SE~\cite{venkatesh2013bridging}}.  Deciding what to include (and exclude) in this paper from this large body of research was challenging and we focused on the \edit{guiding principles}, properties and designs that we hope will support software engineering researchers in applying mixed methods to lead to impactful insights.  Still, applying mixed methods is not a trivial exercise, as we try to demonstrate through the identification of many possible antipatterns. There is a lot more to read and learn about mixed methods, and the references we provide too can provide a starting point for researchers to learn more.

The selecting empirical research methods  paper by Easterbrook et al.~\cite{easterbrook_selecting_2008} notably elaborates on topics such as \textbf{theoretical stance} and the \textbf{role of theory} in quantitative, qualitative and mixed research. The main takeaways from that paper also apply here: a mixed methods paper should have a clear research question that aligns with the research design; the methods used should answer this question; the limitations of any one method can be used to motivate the choice of other methods, but the limitations of and threats associated with mixing should also be reported; and the role of theory and whether the research is inductive or deductive should drive the research design and shape the integrated findings reported. 

We also elaborate on four core \textbf{\edit{guiding} principles} of mixed methods research that are: providing \textit{methodological rationale} for the study design; ensuring the design leads to \textit{novel integrated insights} by the mixing of methods; assuring the rigorous use of individual methods and that they are mixed in a \textit{rigorous} way; and finally, conducting the study in an \textit{ethical} manner.  These principles, at a high level, also apply to any research study (as alluded to above), but we elaborate on how these principles can be considered specifically within the context of a mixed methods research study in SE.  

We further describe the core \textbf{properties} that shape a diverse landscape of mixed methods \textbf{research designs}.  These properties, which include the timing of methods, the inductive or deductive nature of the research, the research design planning or emergence, how data is integrated, and the mixed design, should be carefully considered up front and throughout the research process, and presented clearly in any research manuscript.  We do not provide a comprehensive description of all possible designs, but through several  research scenarios, we show how the principles and properties can be considered as trade-offs and design choices to be made. The integration process throughout the research design is what sets mixed methods apart. 

Our paper further suggests a number of practical considerations. For example, how to \textbf{visualize} the mixed methods research design (as we show for the four scenarios), some potential pitfalls to try to avoid (articulated as \textbf{research antipatterns}), and which limitations to consider and present (guided in part by the principles and how integration should be achieved).  Mixed methods, as mentioned, can be used to offset the limitations of a single method, and can be powerful for investigating socio-technical phenomena in software engineering, but they are challenging to use and often require broad expertise and substantial time commitments from the researchers involved.  

This paper is a first step in understanding how to use mixed methods in software engineering research and we hope that the SE research community will find this guide helpful. 
In closing, we suggest researchers adopt an open and creative stance to research design, and to share back with the community their insights gained not only on software engineering, but also on their research designs as they use mixed methods.\\



 




\section*{Acknowledgements}
We are grateful to Prof. John Grundy for supporting a paper writing workshop at the HumaniSE Lab, Monash University, Melbourne in 2023. We also thank Cassandra Petrachenko and other members of the CHISEL Lab, UVic for feedback on the paper. 
This research was funded in part by the National Science and Engineering Research Council.

\bibliographystyle{spmpsci} 
\bibliography{references.bib}

\section*{Appendix: Glossary} 
\label{appendix}

\paragraph{Data Analysis Techniques}
Researchers specify the methods for analyzing the data collected during the study. These may involve statistical techniques, socio-technical grounded theory (STGT) for data analysis \cite{hoda2024Book}, thematic analysis \cite{braun2012thematic}, or other analytical approaches.   

\paragraph{Data Collection Method}
Researchers must decide how to gather data relevant to their research question (i.e., techniques and methods). This may involve methods such as surveys, interviews, observations, or a combination of these approaches.

\paragraph{Inductive and Deductive}
Inductive and deductive reasoning are two complementary approaches used in research to develop theories, hypotheses, or generalizations. Inductive reasoning involves moving from specific observations or patterns to broader generalizations. 
Deductive reasoning involves moving from general principles or theories to predictions or hypotheses about specific instances. While inductive is commonly associated with \qual~research methods, deductive is frequently associated with \quan~research methods. However, this can vary, as exemplified in our scenarios (Section \ref{sec:scenarios}). 

\paragraph{Meta-inferences}
Meta-inferences refer to making inferences or drawing conclusions about integrating the \qual~and \quan~methods. When integrating \qual~and \quan~data is a central element of the MMR, how the researcher represents this integration and the interpretation of these findings becomes critical. During the MMR discussion, the meta-inferences depicted during this process should be articulated to present how the use of an MMR approach advanced a great understanding of the substantive topic compared to a mono-method approach.

\paragraph{Multi-method study}
A multi-method study may be planned or emergent and involves either two or more \quan~methods or two or more \qual~methods, but not a mix of both \qual~and \quan. The main goal is typically to expand on the insights across methods, and they may be carried out sequentially or in parallel. Distinct research methods can be used in a multi-method design. 

\paragraph{Qualitative and Quantitative}
Qualitative (\qual) and quantitative (\quan) are two primary approaches to research, each with distinct characteristics. 
In this paper, we refer to \qual~data as text and image-based evidence (e.g., words used to describe experiences or stories). In contrast, \quan~data relates to numeric-based evidence (e.g., device analytics or Net Promoter Score). 

\paragraph{Research Design}
A research design involves selecting the most appropriate research design based on the nature of the research question and the desired outcomes (i.e., the researcher's methodological choice if using mono-method, multi-method, or mixed methods).

\paragraph{Research Method}
A research method is a structured approach or systematic process used to investigate, collect, and analyze data to answer questions or solve problems related to software development and engineering. Common research methods include case studies, experiments, surveys, and ethnographic studies, each tailored to explore specific aspects of a research goal.

\paragraph{Research Paradigms} 
\edit{A research paradigm is a framework of beliefs and practices that defines a worldview for understanding phenomena within a discipline, shaping research through ontological, epistemological, and methodological perspectives. Quantitative research, rooted in positivism, seeks objective, measurable realities; qualitative research, aligned with constructivism, explores subjective, socially constructed realities; and mixed methods research, guided by pragmatism, integrates both to address complex questions. Together, these form a trilogy of major research paradigms \cite{Johnson2004}, \cite{Kuhn1977}.}

\paragraph{Research Strategy}
A research strategy refers to the overarching plan or approach that guides the conduct of a research study (e.g., ethnography, socio-technical grounded theory, case study, action research, survey, experiment, etc.). It outlines the steps, methods, and procedures that will be employed to address the research question or objectives effectively.

\end{document}